# The Lookup Technique to Replace Nested-IF Formulas in Spreadsheet Programming

Thomas A. Grossman
University of San Francisco, School of Business & Management, San Francisco CA 94117-1045
tagrossman@usfca.edu

Özgür Özlük
San Francisco State University, College of Business, San Francisco, CA 94132
ozgur@sfsu.edu

Jan Gustavson
Senior Financial Analyst, Visa, Inc., San Francisco, CA 94128-8999

**ABSTRACT**

*Spreadsheet programmers often implement contingent logic using a nested-IF formula even though this technique is difficult to test and audit and is believed to be risky. We interpret the programming of contingent logic in spreadsheets in the context of traditional computer programming. We investigate the "lookup technique" as an alternative to nested-IF formulas, describe its benefits for testing and auditing, and define its limitations. The lookup technique employs four distinct principles: 1) make logical tests visible; 2) make outcomes visible; 3) make logical structure visible; and 4) replace a multi-function nested-IF formula with a single-function lookup formula. It can be used only for certain simple contingent logic. We describe how the principles can be applied in more complex situations, and suggest avenues for further research.*

## 1. INTRODUCTION

Spreadsheets are widely used in business for a variety of important purposes (Coles and Rowley 1996, Ragsdale 2001, Croll 2005, Grossman, Mehrotra and Özlük 2007). A spreadsheet containing cell formulas is a computer program, and all computer programs are susceptible to errors. EuSpRIG maintains a corpus of spreadsheet errors (EuSpRIG 2009). Several scoping-scale research projects (summarized in Panko 2000a, 2000b; Panko 2008) suggest that spreadsheets are particularly vulnerable to errors. However, follow-on research (Powell, Baker and Lawson 2008a, 2008b, and 2008c) indicates that the prevalence and risk of errors in important spreadsheets is not well understood and further research is required.

Spreadsheet programmers often implement contingent logic using the nested-IF formula even though this technique is difficult to test and audit and is believed to be risky. In this paper, we investigate the "lookup technique" for programming contingent logic that can reduce or eliminate the need to use the nested-IF formulas. In certain circumstances, a complex nested-IF can be replaced in a mechanical fashion (without exercise of judgment nor loss of functionality) by a lookup function that is safer to program and provides other benefits.

### 1.1. Nested-IF Formulas Are Considered Risky

Computer programmers often need to program contingent logic. Contingent logic in spreadsheets seems typically to be programmed using a nested-IF formula. We have encountered Nested-If formulas in research observations and informal discussions with experienced spreadsheet programmers. We put out a call to the EuSpRIG list and received many real-world examples of nested-IF formulas.

A nested-IF is created when an IF function used as an argument ("nested") of one or more other IF functions. There is much concern about the use of nested-IF formulas in spreadsheets. Lanza (2006) uses the number of nested-IF formulas as a determinant in the likelihood of risk in a spreadsheet. Croll and Butler (2006) examine the common sources of errors in medical spreadsheet applications and they find the complex nested-IF formulas as one of the main culprits. Thiriez (2004) provides a nested-IF he used that he believes is difficult to audit. Baxter (2005) discusses several spreadsheet auditing tools have been created to address the potential flaws in logic creation which perform checks for logic elements that are known to be particularly error prone (e.g. nested-IF formulas). Spreadsheet audit software such as Spreadsheet Professional and XLAnalyst flag nested-IF formulas as a source of risk.

Although nested-IF formulas may be risky, there is reason to believe that skilled software engineers can manage the risk. As with any challenging programming task, a diligent programmer can write a complex nested-IF accurately. For example, Grossman, Mehrotra and Özlük (2007) found an example of a nested-IF formula with 5 levels of nesting and 11 IF functions that are believed to be correct. However, the prevalence of such diligent programming practices is unknown.

### 1.2. Contribution

This paper considers the "lookup technique" that a programmer in certain circumstances can use instead of a nested-IF formula. The technique uses a lookup function (specifically VLOOKUP or HLOOKUP) to obtain the same functionality as a nested-IF. The general approach has been known for some time. We hypothesize that many spreadsheet programmers have independently discovered it over the years. Read and Batson (2000 p. 6-55) discuss this concept in passing. Jelen (2008) recommends it and focuses on the syntax of the VLOOKUP function.

The contribution of this paper is to present the principles underlying the lookup technique, describe its benefits for testing and auditing, define its limitations, describe how the principles can be applied in more complex situations, and suggest further research.

We interpret the programming of contingent logic in spreadsheets in the context of traditional computer programming. We discuss the problems of nested-IF formulas for spreadsheet testing and spreadsheet auditing. We explain why the lookup approach is beneficial for testing and auditing.

We decompose the lookup method for programming contingent logic into its component parts. We conclude that the lookup approach consists of the application of four distinct principles: 1) make logical tests visible; 2) make outcomes visible; 3) make logical structure visible; and 4) replace a multi-function nested-IF formula with a single-function lookup formula.

We show the limitations of the lookup approach, which is applicable only for nested-IF formulas with certain well-defined characteristics. For those nested-IF formulas that do not admit the use of a lookup function, we believe there are benefits from providing visibility into the logical tests, outcomes, and logical structure. We discuss areas that merit further research.

### 1.3. Structure of Paper

In section 2 we discuss contingent logic in traditional computer programming and point out that nested-IF formulas were once the norm. We describe nested-IF formulas in spreadsheets and discuss the problems posed for testing and auditing. In section 3 we develop an alternative called the "lookup technique", and show how this technique employs four distinct principles. In section 4 we evaluate the lookup technique. We compare it with spreadsheet engineering recommendations, consider resource consumption, and explain when it can and cannot be applied, and discuss the principles of visibility applied to more complex situations, and the challenge of "acceptable errors". We conclude in section 5.

## 2. NESTED-IF FOR SPREADSHEET CONTINGENT LOGIC

In this section we discuss contingent logic, including nested-IF in traditional programming languages and in spreadsheets, and describe the difficulties of testing and auditing a spreadsheet that contains nested-IF formulas.

### 2.1. Contingent Logic in Computer Programming

This paper addresses an important aspect of spreadsheet computer programming, namely the implementation of contingent logic. Contingent logic is required when the programmer wants a computer program to return different values depending on the state of one or more other values.

Traditional programming languages handle contingent logic using programming statements such as If-Then-Else, If-Then-ElseIf, Case, and Switch. Early scientific languages such as FORTRAN 66 provided only the If-Then-Else statement which required cumbersome nesting. Note that the challenge posed by nested-IF predates the spreadsheet!

Later programming languages, such as FORTRAN 77, reduced the need for nested-IF constructs by providing the more powerful If-Then-ElseIf statement. Some modern programming languages also provide the flexible Case and Switch statements.

To better program complex logic, traditional programmers devised conventions such as indenting, line breaks, and color coding. These conventions are universal in traditional programming and are often automated (for example, the Visual Basic editor). However, there are no such standard conventions for programming complex logic in spreadsheets. Indeed, Panko (2008) states that spreadsheet programming "seems to resemble programming practice in the 1950s and 1960s", suggesting there is significant opportunity for improvement in productivity, maintainability, and accuracy. Just as the advent of the If-Then-ElseIf statement took traditional programming out of the 1960's approach of using nested-If statements, we hope that the lookup technique and follow-on research can advance the practice of spreadsheet programming.

## 2.2. Nested-IF in a Spreadsheet

Contingent logic is required when a cell in the program must return different values depending on the value of an input or an intermediate calculation. Contingent logic in spreadsheets is typically programmed using the IF function. The IF function allows the programmer to implement contingent logic that returns one outcome (the "value_if_true" outcome) if the logical test evaluates TRUE, and another outcome (the "value_if_false" outcome) if the logical test evaluates to FALSE. The IF function syntax is as follows:

=IF(logical_test , value_if_true , value_if_false)

When more than one contingency is present, the programmer can nest one IF function inside another. Any of the arguments of the IF function can themselves be functions, providing a high degree of flexibility. For example,

=IF(A3>A4 , IF(A3>A5 , "yes" , "first") , "no")

## 2.3. Difficulty Testing and Auditing Nested-IF

The simple example above is easy to understand. There are but two IF functions, and the outcomes (value_if_true and value_if_false) are text. However, the value_if_true and value_if_false can themselves be IF functions. This nesting process can lead to cell formulas of substantial complexity. Consider this nested-IF formula, drawn from a regularly used spreadsheet[1]:

=IF($D14="Rev",$C14*E$11/$C$11,IF($D14="Units",$C14*E$9/$C$9,IF($D14="MH",$C14*E$12/$C$12,IF($D14="MC",$C14*E$13/$C$13,IF($C14="","","not correct Allocation"))))).

In order to better visualize the formula, we rearrange it and strip out the "$" absolute reference markers:

=IF(B10="Rev",B14*B15/B16,IF(B10="Units",B14*B17/B18,IF(B10="MH",B14*B19/B20,IF(B10="MC",B14*B21/B22,IF(B11="","","not correct Alloc.")))))

We program this nested-If in Figure 1.

---

[1] This nested-IF formula was provided by an informant who indicated the spreadsheet is in routine use.

[Figure 1 spreadsheet showing:]

| | A | B | C | D | E | F | G | H |
|---|---|---|---|---|---|---|---|---|
| 1 | Nested-IF contains: Logical Test Values, Outcome Values, Contingent Logic Structure | | | | | | | |
| 2 | | | | | | | | |
| 3 | | | | | | | | |
| 4 | | | | | | | | |
| 5 | | 4.17 | | | | | | |
| 6 | =IF(B10="Rev",B14*B15/B16,IF(B10="Units",B14*B17/B18,IF(B10="MH",B14*B19/B20,IF(B10="MC",B14*B21/B22,IF(B11="","","not correct Alloc.")))))  | | | | | | | |
| 7 | | | | | | | | |
| 8 | | | | | | | | |
| 9 | Control Inputs | | | | | | | |
| 10 | | MH | | | | | | |
| 11 | | 5 | | | | | | |
| 12 | | | | | | | | |
| 13 | Other Inputs | | | | | | | |
| 14 | | 5 | | | | | | |
| 15 | | 1 | | | | | | |
| 16 | | 2 | | | | | | |
| 17 | | 3 | | | | | | |
| 18 | | 4 | | | | | | |
| 19 | | 5 | | | | | | |
| 20 | | 6 | | | | | | |
| 21 | | 7 | | | | | | |
| 22 | | 8 | | | | | | |
| 23 | | | | | | | | |

Figure 1: Example nested-IF formula

This nested-IF formula poses challenges for a programmer who wants to verify accuracy. One approach to verifying accuracy is to audit the cell formula, by having a programmer (or team of programmers, Panko 2008) inspect the formula. However, it is not a simple exercise to understand the purpose of such a formula. Even after the purpose is understood, it is difficult to verify its correctness.

There are three challenges in auditing this formula. First is the accuracy of the logical test formulas. Second is the accuracy of the outcome formulas themselves; there are six different outcome values in this formula. Each must be separately verified. Third is the accuracy of the contingent logic structure, which is controlled by the location of the five IF functions and the match between the logical tests and the outcomes.

Another approach to verifying accuracy is to use traditional software testing. To test a spreadsheet containing a nested-IF formula, it is necessary to devise test cases that will toggle every possible state of the nested-IF. Such test cases can be challenging or even impossible to discover. Therefore, traditional input-output testing is challenging and might be impossible.

## 3. THE FOUR PRINCIPLES OF THE LOOKUP TECHNIQUE

To ameliorate these problems, programmers sometimes use the lookup technique, which replaces the nested-IF formula with several additional cells and a simple lookup function, as shown in Figure 5 below.

The lookup technique employs four principles. First is to make visible the logical test values (TRUE or FALSE). Second is to make visible the outcome values. Third is to make visible the structure of the contingent logic. Fourth is to use a simple formula to handle the contingent logic. To elucidate the principles underlying the lookup technique, we will take a nested-IF and transform it step-by-step it into a lookup.

### 3.1. First Principle: Make Visible the Logical Test Values

We rewrite our example nested-IF to make visible in the spreadsheet the values of the logical_tests. We replace each logical_test formula with the range name "Test1", "Test2", etc. This yields the following nested-IF formula.

    =IF(Test1,$B$14*$B$15/$B$16,IF(Test2,$B$14*$B$17/$B$18,IF(Test3,$B$14*$B$19/$B$20,
    IF(Test4,$B$14*$B$21/$B$22,IF(Test5,"","Not Correct Alloc.")))))

where

    Test1 replaces    $B$10="Rev"
    Test2 replaces    $B$10="Units"
    Test3 replaces    $B$10="MH"
    Test4 replaces    $B$10="MC"
    Test5 replaces    $B$11=""

We can program this in a spreadsheet as shown in Figure 2.

|   | A | B | C | D | E | F | G | H |
|---|---|---|---|---|---|---|---|---|
| 1 | Nested-IF contains: Outcome Values, Contingent Logic Structure | | | | | | | |
| 2 | Visible in the Spreadsheet: Logical Test Values | | | | | | | |
| 3 | | | | | | | | |
| 4 | | | | | | | | |
| 5 | | 4.17 | | | | | | |
| 6 | =IF(Test1,$B$14*$B$15/$B$16,IF(Test2,$B$14*$B$17/$B$18,IF(Test3,$B$14*$B$19/$B$20,IF(Test4,$B$14*$B$21/$B$22,IF(Test5,"","Not Correct Alloc."))))) | | | | | | | |
| 7 | | | | | | | | |
| 8 | | | | | Logical Tests | | | |
| 9 | | Control Inputs | | Name | Value | Value Formula | | |
| 10 | | MH | | Test1 | FALSE | $B$10="Rev" | | |
| 11 | | 5 | | Test2 | FALSE | $B$10="Units" | | |
| 12 | | | | Test3 | TRUE | $B$10="MH" | | |
| 13 | | Other Inputs | | Test4 | FALSE | $B$10="MC" | | |
| 14 | | 5 | | Test5 | FALSE | $B$11="" | | |
| 15 | | 1 | | | | | | |
| 16 | | 2 | | | | | | |
| 17 | | 3 | | | | | | |
| 18 | | 4 | | | | | | |
| 19 | | 5 | | | | | | |
| 20 | | 6 | | | | | | |
| 21 | | 7 | | | | | | |
| 22 | | 8 | | | | | | |
| 23 | | | | | | | | |

Figure 2: Logical Test Values Made Visible in cells E10:E14

Notice that we added five new formula cells, one for each of the five logical tests. This allows each logical test to be instantly verified for accuracy, whatever the state of the spreadsheet.

### 3.2. Second Principle: Make Visible the Outcome Values

We next rewrite our example nested-IF to make visible in the spreadsheet the values of the Outcomes. We replace each outcome (formula or text) with the range names "Value1", "Value2" and so forth. This yields the following nested-IF formula.

=IF(Test1,Value1, IF(Test2,Value2, IF(Test3,Value3, IF(Test4,Value4, IF(Test5,Value5,Value6)))))

where Value1 through Value6 is defined as above and

Value1 replaces   $B$14*$B$15/$B$16
Value2 replaces   $B$14*$B$17/$B$18
Value3 replaces   $B$14*$B$19/$B$20
Value4 replaces $B$14*$B$21/$B$22
Value5 replaces   ""
Value6 replaces  "not correct Allocation"

We can program this in a spreadsheet as shown in Figure 3:

Figure 3: Outcome Values Made Visible in cells E18:23

Notice that we added six new formulas cells, one for each of the six outcomes. This allows each outcome to be instantly verified for accuracy, whatever the state of the spreadsheet. If we intend to perform traditional testing, these outcome values can be tested for every test case, not just for the test cases that cause the contingent logic to display that outcome.

### 3.3. Third Principle: Make Visible the Structure of the Contingent Logic

Our next improvement is to modify the spreadsheet to make the structure of the contingent logic more visible. We rearrange the cells so that the logical tests are placed next to their corresponding outcome values.

Figure 4: Structure of Contingent Logic Made Visible. The nested-IF in cell B5 displays the value in column F corresponding to the first appearance of TRUE in Column C, or the value in row 17 if none are TRUE. Note that the nested-IF contains the contingent logic, and the contingent logic is also visible in the spreadsheet.

Notice in Figure 4 that if Test1 is TRUE, then the value in the same row is returned. If Test1 is FALSE and Test2 is TRUE, the value in the same row as Test2 is returned, and so forth. If all five logical tests are FALSE, the final "(otherwise)" value is returned.

### 3.4. Fourth Principle: Simplify the Formula by Replacing Nested-IF with a Lookup

Our next improvement is to replace the complex, multi-function nested-IF cell formula with a simple, single-function VLOOKUP cell formula. The new formula is

VLOOKUP(TRUE,E12:F17,2,FALSE)

Notice that we must type the logical test value TRUE in cell E17 to cause the VLOOKUP to return the last Outcome value. Figure 5 presents a spreadsheet with the VLOOKUP formula and the equivalent nested-IF formula.

| | A | B | C | D | E | F | G | H |
|---|---|---|---|---|---|---|---|---|
| 1 | Simpler Formula: VLOOKUP instead of Nested-IF | | | | | | | |
| 2 | | | | | | | | |
| 3 | | 4.17 | VLOOKUP FORMULA | | | | | |
| 4 | =VLOOKUP(TRUE,E12:F17,2,FALSE) | | | | | | | |
| 5 | | | | | | | | |
| 6 | | 4.17 | Nested-IF FORMULA (for comparison) | | | | | |
| 7 | =IF(Test1,Value1,IF(Test2,Value2,IF(Test3,Value3,IF(Test4,Value4,IF(Test5,Value5,Value6))))) | | | | | | | |
| 8 | | | | | | | | |
| 9 | | Control Inputs | | | Contingent Logic Structure | | | |
| 10 | | MH | | | Logical Tests | | Outcomes | |
| 11 | | 5 | | | Name | Value | Value | Name |
| 12 | | | | | Test1 | FALSE | 2.500 | Value1 |
| 13 | | Other Inputs | | | Test2 | FALSE | 3.750 | Value2 |
| 14 | | 5 | | | Test3 | TRUE | 4.167 | Value3 |
| 15 | | 1 | | | Test4 | FALSE | 4.375 | Value4 |
| 16 | | 2 | | | Test5 | FALSE | | Value5 |
| 17 | | 3 | | | | TRUE | Not Correct Alloc | Value6 |
| 18 | | 4 | | | | | | |
| 19 | | 5 | | | Note: Cell E19 contains the logical value TRUE. | | | |
| 20 | | 6 | | | This is necessary for the VLOOKUP in the case | | | |
| 21 | | 7 | | | where Test1 through Test5 have Value FALSE. | | | |
| 22 | | 8 | | | | | | |
| 23 | | | | | | | | |

Figure 5: Lookup Technique for Contingent Logic. The VLOOKUP in Cell B80 cell address displays the value in column D corresponding to the first appearance of TRUE in Column C.

Compared to Figure 4, Figure 5 replaces the complex nested-IF formula with a simpler VLOOKUP that is much easier to verify.

Compared to Figure 1, in Figure 5 the single nested-IF cell has been replaced by 12 cells. Five cells report the value of the five logical tests. Six cells report the value of the six outcomes. One cell contains the structure (and only the structure) of the contingent logic.

### 4. EVALUATION AND FURTHER RESEARCH

The lookup technique for contingent logic takes a complex nested-IF that is difficult to understand, test, and audit and replaces it with new cells that are easy to understand, are testable, and are easier to audit.

### 4.1. Comparison with Spreadsheet Engineering Recommendations

The lookup technique for contingent logic expands a single cell with a complex formula into many cells with simple formulas. It is consistent with the programming advice to "write for the reader". This is well-known in the traditional programming world, and is recommended by Powell and Baker textbook (2007b, page 101) . In contrast, Raffensperger (2003) recommends using the fewest cells possible. This recommendation is driven by the desire to present compact reports; this is easily handled by placing reports on a separate worksheet that references the computational results.

### 4.2. Resource Consumption: Cells and Computation Time

The lookup technique for contingent logic increases the number of cells required by moving "covert computations" out of the nested-IF and into their own cells. If the number of cells available is a constraint, then this is problematic. For example, Thiriez (2004) describes a model that uses all 32,000 rows of an Excel 2003 spreadsheet. With the availability of one million rows in Excel 2007, the availability of cells should be an issue rarely if ever.

The lookup technique could increase computation time because evaluations of logical tests and values that were previously computed only when needed by the contingent logic are now always evaluated. However, with modern computer hardware this issue will likely only rarely be a problem.

**4.3. Branch-on-False**

The nested-IF formulas in this paper have a pattern to the structure of the logic that we call "branch-on-False". In the upper-level IF functions the value_if_true is an outcome and the value_if_false is a nested-IF:

    IF(logical_test , value , IF(…)).

The lookup technique we describe works directly only for branch-on-false. This limitation should be explored by further research. First, with simple modification to our approach it should be easy to make the lookup technique work for "branch-on-true", which is of the form IF(logical_test , IF(…) , value). Second, by selectively reversing the logical tests it should be possible to convert a nested-IF that has a mix of branch-on-true and branch-on-false into a nested-IF that has only branch-on-false, thereby admitting the use of the lookup technique.

**4.4. Complexity of Logical Branching**

The act of making the structure of the logic visible is difficult in the situation where there is an IF function that has further IF functions for both value argument, which has the following form:

    IF(logical_test , IF(…) , IF(…)).

It is not possible to directly apply the lookup technique on this type of complex contingent logic. Therefore, such nested-IFs do not admit the lookup approach. This topic merits further research; what can we do to enhance the programming of complex contingent logic?

**4.5. Principles of Visibility Applied to Complex Branching**

With complex branching, it is not possible to replace a multi-function nested-IF formula with a single-function lookup formula. Whether there is an option to make an improvement merits further research. Regardless, the "visibility" benefits seem to be available. These include making logical tests visible; making outcomes visible; making logical structure visible. Further research should explore how to apply these three principles to complex branching.

**4.6. Acceptable Errors**

In some nested-IF formulas we have seen, outcome values can evaluate to an error when the outcome value is not returned by the nested-IF. An example is a divide-by-zero error that is trapped by the nested-IF. If the programmer applies the principle of making outcome values visible, such error values will now appear in the spreadsheet. If it is unacceptable to have error values in the spreadsheet, then the error message might need to be detected and flagged in some fashion. This topic merits further research.

**5. CONCLUSION AND FUTURE DIRECTIONS**

We discuss the "lookup technique" for programming contingent logic in spreadsheets. We illustrate a step-by-step process to convert a simple nested-IF into a lookup. We explain the benefits in terms of the lookup technique in terms of testing and auditing.

We identify four principles associated with the technique. We propose that these four principles can and should be used during spreadsheet design. If they are implemented during design and programming of a spreadsheet, we hypothesize that they will reduce the total effort to create the spreadsheet (especially when one considers how long it takes to code a complex nested-IF formula), and reduce the time to debug the spreadsheet. Such hypotheses merit empirical evaluation, and we believe deployment of this approach to programmers to be an important avenue of future research.

We show that the lookup technique does not apply to certain "complex" contingent logic situations that present as a complex nested-IF formula. It would be desirable to determine what particular types of spreadsheets benefit from the proposed approach and what types may not. We anticipate that some of the principles can still be applied and yield benefits. We illustrated the lookup technique in a "spreadsheet remediation" situation, by converting an existing nested-IF into a better lookup technique.

 

The lookup technique yields code that in our opinion is more readable because individual cell formulas are shorter. We further believe that the task of verifying or auditing the accuracy of contingent logic is easier with the lookup approach than with the nested-IF approach. These assertions merit empirical testing.

As with any programming principles, there will always be times when a programmer chooses not to adhere to them. The four principles and the use of the lookup technique in general, should be viewed as guidelines rather than rigid rules.

We view the use of the lookup technique, and any future more advanced techniques for programming contingent logic in spreadsheets to be comparable to the traditional programming norm of systematic indentation of a complex series of statements. It is interesting to consider whether other such spreadsheet norms could be developed. Similar to traditional programming languages, we would like to see research to develop the spreadsheet equivalent of a Case statement.

Finally, a referee drew our attention to the EUSES corpus[2] as a repository of spreadsheets. We hope to use this corpus in future research efforts.

**ACKNOWLEDGEMENTS**

We would like to thank the EuSpRIG program chair and the anonymous referees for their constructive feedback. Any errors of omission or commission are the responsibility of the authors.

---

[2] **http://esquared.unl.edu/wikka.php?wakka=EUSESSpreadsheetCorpus**